\documentstyle[aasms4,12pt]{article}
\begin{document}

\def\P{\bar{\Phi}}
\def\st{\sigma_{\rm T}}
\def\vk{v_{\rm K}}
\def\sles{\lower2pt\hbox{$\buildrel {\scriptstyle <}
   \over {\scriptstyle\sim}$}}
\def\sgreat{\lower2pt\hbox{$\buildrel {\scriptstyle >}
   \over {\scriptstyle\sim}$}}

\title{The Axisymmetric Pulsar Magnetosphere}

\author{Ioannis Contopoulos \altaffilmark{1},
Demosthenes Kazanas \altaffilmark{2} and 
Christian Fendt \altaffilmark{3} }
\affil{}
\altaffiltext{1}{Physics Department, University of
Crete, P. O. Box 2208, Heraklion 71003, Greece}
\altaffiltext{2}{NASA/Goddard Space Flight Center, Greenbelt, MD 20771}
\altaffiltext{3}{Astrophysikalisches Institut Potsdam, D-14482 Potsdam, 
Germany}

\begin{abstract}

We present, for the first time, the structure of the axisymmetric
force--free magnetosphere of an aligned rotating magnetic dipole, in
the case in which there exists a sufficiently large charge density
(whose origin we do not question) to satisfy the ideal MHD condition,
${\bf E\cdot B}=0$, everywhere.  The unique distribution of electric
current along the open magnetic field lines which is required for the
solution to be continuous and smooth is obtained numerically.  With
the geometry of the field lines thus determined we compute the
dynamics of the associated MHD wind.  The main result is that the
relativistic outflow contained in the magnetosphere is not accelerated
to the extremely relativistic energies required for the flow to
generate gamma rays.  We expect that our solution will be useful as
the starting point for detailed studies of pulsar magnetospheres under
more general conditions, namely when either the force-free and/or the
ideal MHD condition ${\bf E\cdot B}=0$ are not valid in the entire
magnetosphere.  Based on our solution, we consider that the most
likely positions of such an occurrence are the polar cap, the
crossings of the zero space charge surface by open field lines, and
the return current boundary, but {\em not} the light cylinder.

\keywords{magnetic fields --- MHD --- pulsars: general} 

\end{abstract}
\baselineskip = 15pt
\section{Introduction}

The issue of the structure of neutron star magnetospheres, roughly
thirty years from the seminal paper of Goldreich \& Julian (1969,
hereafter GJ) which outlined its basic physics, remains still
open. The discovery of radiation emission from radio to gamma-rays
with the pulsar period has motivated the modification of the original
GJ model to include features, such as charge gaps, which would lead
to the acceleration of particles necessary to produce the observed
radiation.

Thus, the ubiquitous presence of high energy radiation from
pulsars, in agreement with simple scaling laws (Harding 1981) which
make no particular demands on the magnetic field structure, has
prompted a number of authors to suggest that the gamma-ray emission
results from an unavoidable violation of the `assumption of strictly
dissipation--free flow would lead to singularities in both the flow
and the magnetic field, occuring a short distance beyond the
light--cylinder' (Shibata~1995; Mestel \& Shibata~1994). Such
suggestions were corroborated by solutions in which the field lines
appeared exhibiting kinks and discontinuities which have been
presented occasionally in the literature (e.g. Michel~1982; Sulkanen
\& Lovelace~1990).  Based on the conviction that such singularities
might simply reflect the shortcomings of our numerical methods and not
the physical path nature chooses, we have decided to investigate the
issue ourselves.  Our original hope was that an exact solution of the
magnetohydrodynamic (MHD) structure of a pulsar magnetosphere would
provide also the detailed dynamics of the acceleration of the
associated MHD wind which could then resolve the issue of high energy
emission from these objects.

As often done in the literature, we will work in the framework of
ideal (dissipation--free) magnetohydrodynamics, and will assume that
dissipation effects are minor perturbations to the global picture that
we will present. We do acknowledge that this idealized picture will
not be valid in the presence of physical magnetospheric instabilities
(see discussion below), but assuming for the moment that such
instabilities are absent, we can justify our approach as follows
(e.g. Bogovalov~1997): The gravitational field of an unmagnetized
neutron star would reduce a thermally supported atmosphere to a height
of a few centimeters and the star would be surrounded by an effective
vacuum.  Real stars, however, are magnetized and rotation generates
huge electric fields which, not only pull charged particles from the
surface, but also accelerate them in curved trajectories, generating
thus curvature photons which produce electron--positron pairs in the
curved magnetic field. An electromagnetic cascade develops within a
few stellar radii, and the density of plasma increases up to the point
that it becomes high enough to screen the electric field component
parallel to the magnetic field.  This mechanism is obviously
self--regulating. Wherever the density of plasma drops below the
absolute minimum required to short--out the parallel component of the
electric field, `gaps' will develop, inside which ${\bf E\cdot B} \neq 0$,
and the electromagnetic cascade mechanism will regenerate the missing
charges.\footnote{Electric charges are needed to provide, not only the
space charge density, but also the electric current required in a
magnetosphere where the parallel component of the electric field is
self--consistently shorted--out everywhere.}

The condition that ${\bf E\cdot B}=0$ is precisely the condition of
ideal MHD, and it will be a valid approximation in the study of the
magnetosphere over length scales much larger than the size of the
various `gaps'.  We will further assume that the main stresses in the
magnetosphere are magnetic and electric (with inertial and
gravitational ones being negligible, at least near the surface of the
star), and a valid magnetospheric model will be one where
\begin{equation}
\frac{1}{c}{\bf J\times B} + \rho_e {\bf E} =0\ .
\label{JBE}
\end{equation}
Here, ${\bf J}$, ${\bf B}$, and ${\bf E}$ are the electric current
density, magnetic and electric fields respectively; and
$\rho_e=\nabla\cdot{\bf E}/(4\pi)$ is the electric charge density in
the magnetosphere. We need to note that this simplifying working
assumption is not valid in the neutron star interior (on to which the
magnetic field is anchored), and in singular magnetospheric regions
(current sheets subject to non--vanishing Maxwell stresses which must
be balanced by a thermal pressure).  As we will see, we do not have to
worry about these regions as long as they are self--consistently
treated as boundaries in our problem. More problematic, however, is
the force--free assumption, namely that the inertial effects
associated with the plasma outlflow are generally small. This
assumption may actually be valid in the case of relatively fast
rotating, young neutron stars, as in this case the luminosity
associated with the gamma ray emission is only a small fraction of the
power associated with the spin down of the neutron star through the
magnetic torques; (Arons~1981; Daugherty \& Harding~1982; Cheng, Ho \&
Ruderman~1986); however, for older, slower rotating neutron stars for
which the radiative efficiency gets close to 1 (Harding 1981), we
expect this assumption to be violated.  Since little progress has been
done in the study of the general problem when matter inertia is not
negligible (e.g. Camenzind~1986), we have decided to study the
force--free case, bearing in mind that our solution can only serve as
a first approximation to the physical one.

Having presented our assumptions and their limitations, let us now
investigate more closely what eq.~(\ref{JBE}) and the assumption of
ideal MHD imply about the overall structure of the rotating neutron
star magnetosphere.

\section{The pulsar equation}

A convenient approach in steady--state axisymmetric MHD is to work
with the flux function $\Psi$ defined through
\begin{equation}
{\bf B_p}=\frac{\nabla\Psi \times \hat{\phi}}{R}\ ,
\end{equation}
where, $\bf B_p$ is the poloidal ($R,Z$) component of the magnetic
field in a cylindrical coordinate system ($R,\phi,Z$).  Magnetic field
lines lie along magnetic flux surfaces of constant $\Psi$.  At each
point, $\Psi$ is proportional to the total magnetic flux contained
inside that point; it is also related to the $\phi$ component of the
vector potential.  Ideal force--free MHD requires that
\begin{equation}
B_\phi=\frac{A(\Psi)}{R}\ ,
\label{Bphi}
\end{equation}
where $A(\Psi)$ is a yet to be determined function.  Obviously, the
poloidal electric current $I\equiv cA/2$ is also a function of $\Psi$,
which means that poloidal electric currents are required to flow along
(and not across) flux surfaces.\footnote{It is implicitly assumed here
that, in order for the electric circuit to close, the outgoing current
is able to cross flux surfaces and return to the star in some domain
which is sufficiently remote for our ideal MHD approach to be
applicable (see, however, also \S~6).}  Note that this {\em does not}
in general imply that the electric currents flow along magnetic field
lines, since the electric term in eq.~(\ref{JBE}) cannot in general be
neglected. In fact, using eq.~(\ref{E}) (below), one can see that
the currents are a sum of a current along the total field plus the
convection current due to corotation of the space charge density.
Finally, the electric field is given by
\begin{equation}
{\bf E}=\frac{R\Omega}{c}{\bf B_p}\times \hat{\phi}\ ,
\label{E}
\end{equation}
and is clearly perpendicular to ${\bf B}$. $\Omega$ is the angular
velocity of rotation of the neutron star on to which the magnetosphere
is anchored, and can directly be thought of as the angular velocity of
{\em rigid} rotation of the magnetic field lines\footnote{In a non
force--free treatment where relativistic acceleration near the star is
self--consistently taken into account, the angular velocity of
isorotation of magnetic field lines is slightly smaller than $\Omega$
(e.g. Mestel \& Shibata~1994). As is often done in the literature, we
simplify the problem by assuming the two are equal.}  ({\em not} of
the magnetospheric plasma!).  Eq.~(\ref{JBE}) can now be written in
the equivalent form
\begin{equation}
(1-x^2)
\left(\frac{\partial^2\Psi}{\partial x^2}
-\frac{1}{x}\frac{\partial\Psi}{\partial x}
+\frac{\partial^2\Psi}{\partial z^2}\right)
-2x\frac{\partial\Psi}{\partial x}=-R_{LC}^2
AA'\ ,
\label{pulsareq}
\end{equation}
where we have introduced the convenient notation $x\equiv R/R_{LC}$
and $z\equiv Z/R_{LC}$, with $R_{LC}\equiv c/\Omega$ the distance from
the axis where a particle corotating with the star would rotate at the
speed of light (the so called `light cylinder'); and $(\ldots)'\equiv
{\rm d}/{\rm d}\Psi$. Eq.~(\ref{pulsareq}) is the well known pulsar
equation (Michel~1973a; Scharlemann \& Wagoner~1973). Solutions to
this equation have been found for specific functional
forms of the current distribution $A(\Psi)$. Michel has presented
solutions for $A = $ const. and $A = - 2 \Psi$ for which this equation
becomes linear and the usual techniques can be applied to derive the
form of the field geometry for $x \le 1$. Michel has also presented
solutions for a quadratic form of $A(\Psi)$ corresponding to a
magnetic monopole solution.

One can check that, when this equation is satisfied, the space charge
(or GJ charge) density is conveniently given by
\begin{equation}
\rho_e=\left(\frac{\Omega}{4\pi c}\right)
\frac{-2B_z+AA'}{1-x^2}\ .
\label{rhoe}
\end{equation}
Eq.~(\ref{pulsareq}) is elliptic, and according to the theory of
elliptic equations (albeit the linear ones), the solution at all $x$
and $z$ is uniquely determined when one specifies the values of either
$\Psi$ (Dirichlet boundary conditions) or the normal derivative of
$\Psi$ (Neumann boundary conditions) along the boundaries, i.e. along
the axis $x=0$, the equatorial plane $z=0$, and infinity (as one
expects, and as we will see in practice, the boundary conditions at
infinity will not affect the solution near the origin and the light
cylinder).  Unfortunately, this procedure does not work since
eq.~(\ref{pulsareq}) is singular at the position of the light cylinder
$x=1$.  The singularity at $x=1$ imposes the important constraint that
\begin{equation}
\frac{\partial\Psi}{\partial x}=\frac{1}{2}AA'\ ,
\label{singular}
\end{equation} 
at all points along the light cylinder, and as a result, not all
distributions of electric current along the open field lines
$A=A(\Psi)$ will lead to solutions which cross the light cylinder
without kinks or discontinuities. In fact, as we will see, {\em there 
exists  a unique (?) distribution $A=A(\Psi)$ which allows for the 
continuous and smooth crossing of the light cylinder}.

How can one determine what is {\em the} distribution of $A=A(\Psi)$?
One sees directly that eq.~(\ref{singular}) has precisely the form of
a boundary condition along the light cylinder which will allow for the
solution of eq.~(\ref{pulsareq}) {\em inside} and {\em outside} the
light cylinder. In other words, eq.~(\ref{singular}) determines the
normal derivative of $\Psi$ along the light cylinder, when $A=A(\Psi)$
is known, which can be used to solve the original elliptic equation
both inside and outside $x=1$. The inside solution will yield the
distribution of $\Psi(x=1^-,z)$, the outside solution the distribution
of $\Psi(x=1^+,z)$, and in general, $\Psi(x=1^-,z)$ will not be equal
to $ \Psi(x=1^+,z)$, unless of course one is `lucky enough' to try
{\em the} correct distribution of $A(\Psi)$. To the best of our 
knowledge, all to date efforts to
solve the distorted dipole magnetosphere lead to the conclusion that
`the assumption of strictly dissipation--free flow would lead to
singularities in the magnetic field, occuring a short distance beyond
the light--cylinder' (Mestel \& Shibata~1994).  Several unsuccessful
attempts to solve eq.~(\ref{pulsareq}) in all space, have concluded
in favor of the `inevitability of the break--down of continuity and 
smoothness' of these solutions.

On the other hand, motivated by the fact that the singularity at the
light cylinder is none other than the relativistic unique
generalization of the familiar Alfv\'{e}n point of the
non--relativistic theory (Li \& Melrose~1994; Camenzind~1986) under
force--free conditions, we were more optimistic in that such a
continuous and smooth solution actually exists; we considered the fact
that it had not been found not a sufficient argument against its
existence. We are all too familiar with the difficulties associated
with the continuous and smooth crossing of the non--relativistic
Alfv\'{e}n singular point (Contopoulos 1994, 1995) and no one believes
anymore early solutions which are either discontinuous or show kinks
at or around the position of the Alfv\'{e}n point.  Moreover, smooth
and continuous solutions of the special relativistic problem exist in
a couple of idealized cases: the simple monopole (Michel~1991), a
collimated proto stellar jet (Fendt, Camenzind, Appl~1995), and a
differentially rotating self--similar disk--wind (Li, Chiueh \&
Begelman~1992; Contopoulos~1994).  We really cannot see why would the
relativistic Alfv\'{e}n point of the rotating dipole problem fare
otherwise.

\section{The Numerical Procedure}

We have, therefore, set out to obtain {\em the} solution of the
force--free rotating relativistic dipole problem in all space, without
kinks or discontinuities on or around the light cylinder, 
using the following relaxation type technique:
\begin{enumerate}
\item We chose some (any) initial trial electric current distribution
(the one which corresponds to the relativistic monopole solution
yields easier numerical convergence).
\item We then solve the problem both inside and outside the light cylinder,
and thus obtain the two distributions $\Psi(x=1^-,z)$ and
$\Psi(x=1^+,z)$ along the light cylinder, which in general differ.
\item We correct for the distribution of
$A(\Psi)$ along field lines which cross the light cylinder as
follows: At each point $z$ of the light cylinder correspond values of
$AA'(\Psi(x=1^-,z))$ and $AA'(\Psi(x=1^+,z))$. From them, we obtain
the new distribution
\begin{eqnarray}
AA'|_{\rm new}(\Psi) & = &
\mu_1 AA'(\Psi(x=1^-,z))+\mu_2 AA'(\Psi(x=1^+,z)) \nonumber \\
 & & + \mu_3[\Psi(x=1^+,z)-\Psi(x=1^-,z)]\ ,
\label{correction}
\end{eqnarray}
\begin{equation}
{\rm for}\ \Psi=\frac{1}{2}[\Psi(x=1^-,z)+\Psi(x=1^+,z)]\ ,
\label{newPsi}
\end{equation}
with weight factors $\mu_1+\mu_2=1$, and $\mu_3\ll 1$ chosen
empirically in order to facilitate numerical
convergence. Eq.~(\ref{correction}) can be seen as an equation
$AA'|_{\rm new}=AA'|_{\rm new}(\Psi)$, which is solved to yield the
distribution $A|_{\rm new}(\Psi)$, with the boundary condition
$A|_{\rm new}(\Psi=0)=0$.  Obviously we also require that $A=0$ along
closed field lines.
\item We repeat steps 1 to 3 until the difference 
$[\Psi(x=1^+,z)-\Psi(x=1^-,z)]$ becomes numerically negligible along
all points of the light cylinder.
\end{enumerate}
Obviously, at that point, the solution will be both continuous
(i.e. $\Psi(x=1^-,z)=\Psi(x=1^+,z)$), and smooth
(i.e. $\partial\Psi/\partial x|_{x=1^-}=
\partial\Psi/\partial x|_{x=1^+}=AA'|_{\Psi(x=1,z)}/2$) at the light
cylinder.

In order to be able to solve the problem all the way to infinity, we
have rescaled our $x$ and $z$ variables. Inside the light cylinder, we
rewrite eq.~(\ref{pulsareq}) in the variables $X_{\rm in}=x$, $Z_{\rm
in}=z/(1+z)$, whereas outside the light cylinder, we work in the
variables $X_{\rm out}=(x-1)/x$, $Z_{\rm out}=z/(1+z)$. In both cases,
our computational domain becomes
\begin{equation}
0\leq X_{{\rm in},\ {\rm out}} \leq 1\ ,\ {\rm and}\ 0\leq 
Z_{{\rm in},\ {\rm out}} \leq 1\ ,
\end{equation}
with the appropriate boundary conditions 
\begin{equation}
\Psi(X_{\rm in}=0,Z_{\rm in})=0\ ,
\nonumber
\end{equation}
\begin{equation}
\Psi(X_{\rm in},Z_{\rm in}\ {\rm around\ the\ origin})=
\Psi_{\rm dipole}=\left(\frac{m}{R_{LC}}\right)
\frac{x^2}{(x^2+z^2)^{3/2}}\ ,
\nonumber
\end{equation}
\begin{equation}
\left.\frac{\partial\Psi}{\partial Z_{\rm in}}
\right|_{X_{\rm in},Z_{\rm in}=0}=0\ ,\ 
\left.\frac{\partial\Psi}{\partial X_{\rm in}}
\right|_{X_{\rm in}=1,Z_{\rm in}}
=\left.\frac{AA'}{2}\right|_{\Psi(X_{\rm in}=1,Z_{\rm in})}\ ,
\end{equation}
inside the light cylinder, and
\begin{equation}
\left.\frac{\partial\Psi}{\partial X_{\rm out}}
\right|_{X_{\rm out}=0,Z_{\rm out}}
=\left.\frac{AA'}{2}\right|_{\Psi(X_{\rm out}=0,Z_{\rm out})}\ ,
\Psi(X_{\rm out},Z_{\rm out}=0)=\Psi_{\rm open}\ ,
\end{equation}
outside. Here, $\Psi_{\rm open}$ is the value of $\Psi(X_{\rm
in}=1,Z_{\rm in}=0)$ obtained from the solution inside the light
cylinder which is obtained first and $m$ is the dipole magnetic
moment.  We chose $\Psi(X_{{\rm in},\ {\rm out}},Z_{{\rm in},\ {\rm
out}}=1)=0$ and $\partial\Psi/\partial Z_{\rm out}(X_{\rm
out}=1,Z_{\rm out})=0$ at the upper and outer right boundary
respectively.  As expected, the solution around the origin is indeed
independent of the exact choice of these boundary conditions. In fact,
it assumes a monopole type (i.e. radial) structure at large distances.

We have used the elliptic solver provided in Numerical Recipes (Press
{\em et al.}~1988).  This implements the simultaneous over relaxation
solution of a linear (i.e. with coefficients fixed in space) elliptic
equation with Chebyshev acceleration, in a rectangular computational
grid.  Our present problem is clearly non--linear. One can, however,
regard $AA'(\Psi(x,z))$ as a function of position at each iteration
step, after making an initial guess for $\Psi(x,z)$. The iteration
procedure is repeated until the input and output $\Psi$'s are the
same.  In a grid of $40\times 40$ points, we reach convergence in
about 1000 iterations.  As a test of our numerical code, we checked
three known solutions which are reproduced with great accuracy in
figure~1: the rotating relativistic dipole with zero poloidal current
(only inside the light cylinder; Michel~1973b), the rotating
relativistic monopole (Michel~1973a), and the solution in Michel~1982.
As we said, we generate the distributions $\Psi=\Psi(x,z)$ which solve
eq.~(\ref{pulsareq}) inside and outside the light cylinder, and then
correct the electric current distribution to one which will
(hopefully) lower the differences $\Psi(x=1^+,z)-\Psi(x=1^-,z)$ along
the light cylinder.  The reader can get an idea of the discontinuities
that the electric current distribution correction iteration goes
through in fig.~2. The solution is {\em extremely} sensitive to the
electric current distribution, and small deviations from the correct
current distribution reflect to large kinks/discontinuities at the
light cylinder. In view of this sensitivity of the solution to the 
current distribution it becomes apparent that a simple guess of its
form is likely to result in discontinuities in the solutions.

\section{The Solution}

The procedure described in the previous section is repeated 50 times,
at which point we obtain a magnetospheric structure which is
sufficiently smooth and continuous around the light cylinder
(figure~3). The last open field line (thick line) corresponds to
\begin{equation}
\Psi_{\rm open}=1.36 \Psi_{\rm pc}\ ,
\end{equation}
where, $\Psi_{\rm pc}\equiv m/R_{LC}$ corresponds to the last field
line which closes inside the distance to the light cylinder in the
nonrelativistic dipole solution. As expected from our intuition based
on the current--free distorted dipole solution, $\Psi_{\rm open}>
\Psi_{\rm pc}$, and contrary to one's naive expectation,
the present magnetically dominated system does not reach a closed field
line structure outside the light cylinder, but rather opts (as we will
see) for a quasi--radial structure.  A nice physical way to see this
effect is that the equivalent `weight' associated with the
electromagnetic field energy pulls the lines open because of the
magnetospheric rotation (Bogovalov~1997).

The main electric current (which, for an aligned rotator, flows into
the star) is equal to 
\begin{equation}
I= 0.6 I_{\rm GJ}\ ,
\end{equation}
where, $I_{\rm GJ}\equiv \Omega^2 m/c$ is the electric current one
obtains by assuming that electrons (positrons in a counter aligned
rotator) with GJ number density stream outward at the speed of light
from the nonrelativistic dipole polar cap.  This electric current is
distributed along the inner open field lines $0<\Psi<1.08\Psi_{\rm
pc}$, as seen in figure~4.  The electric current distribution is close
to the one which corresponds to a rotating monopole with the same
amount of open field lines (dashed line), but varies slightly, in
particular in that a small amount of return current ($I_{\rm return}=
0.03 I_{\rm GJ}$) flows in the outer $1.08\Psi_{\rm pc} < \Psi <
1.36\Psi_{\rm pc}$ (the bulk of the return current obviously flows
along the boundary between open and closed lines, and along the
equator, i.e. the thick line in figure~3). This is very interesting in
view of the fact that the equivalent monopole current distribution
comes close to generate a continuous solution, although the physical
behavior of the inside and outside solutions differ near the light
cylinder (figure~1c; Michel~1982). We would like to emphasize that
several trials of this procedure with different initial current
distributions have all converged to the same final distribution shown
in figure~4. This suggests that there may in fact exist a {\em unique
poloidal electric current distribution consistent with the assumptions
of our treatment}.

We would like to give particular emphasis to a subtle point in our
numerical treatment of the interface between the open and closed field
lines within the light-cylinder.  The numerical relaxation procedure
determines $AA'(\Psi)$, and $A(\Psi)$ is obtained by integrating $AA'$
from $\Psi=0$ to $\Psi_{\rm open}$. This implies that {\em there is no
a priori guarantee} that $A(\Psi_{\rm open})$ be equal to zero, and in
fact it is not. The reader can convince him/herself that, because of
north--south symmetry, this implies that a return current sheet equal
to $-A(\Psi_{\rm open})$ flows along the equator and along the
interface between open and closed field lines.  Since no poloidal
electric current can flow inside the closed domain, there is an
unavoidable discontinuity in $B_\phi$ across the interface, and this
can only be balanced by a similar discontinuity in $B_p$!  This effect
is {\em numerically entirely missed} if one naively considers the
expression for $AA'$ as given in fig.~4, where $AA'\rightarrow 0$ for
$\Psi\rightarrow \Psi_{\rm open}$, since one will then be missing the
delta function (not shown in fig.~4) which corresponds to the step
discontinuity in $A$ (e.g. Michel~1982).  A finite resolution
numerical grid will not discern an infinite jump in $A(\Psi)$, and
therefore, we treat this problem by artificially transforming the step
discontinuity into a smooth (Gaussian) transition in $A$ over an
interval $0.1 \Psi_{\rm open}$. We note that a similar problem does
not arise in the split monopole case, since the current sheet there
extends all the way to the origin, and can be simply treated as an
equatorial boundary.

The null line, i.e. the line with zero GJ space charge is shown
dotted. The crossings of the null line by open field lines have often
been suspected to be the regions where pulsar emission originates
(Cheng, Ho \& Ruderman~1986; Romani~1996).  We plan to investigate the
detailed microphysics of the gaps that will appear around these
regions in a forthcoming publication (see also
\S~6).  According to eq.~(\ref{rhoe}), at large distances, the null 
line asymptotically approaches the field line $\Psi=1.08\Psi_{\rm pc}$
along which $AA'=0$. Well within the light cylinder, the null line is
simply given by the locus of points where the condition ${\bf
\Omega\cdot B}=0$ (or equivalently $B_z=0$) is satisfied.

Knowing {\em the} poloidal electric current distribution along the
open magnetic field lines, we can also derive the asymptotic structure
of our solution at distances $x\gg 1$.  One can directly see that the
flux function becomes then a function of the angle $\theta$ from the
axis of symmetry, and consequently, the poloidal field lines will be
radial.  The distribution of $\Psi$ with angle $\theta$ can be
obtained through the numerical integration of the equivalent form of
the pulsar equation for asymptotically radial field lines,
\begin{equation}
\frac{{\rm d}^2\Psi}{{\rm d}t^2}=-\frac{{\rm d}\Psi}{{\rm d}t}
\frac{1+2t^2}{t(1+t^2)}+\frac{R_{LC}^2AA'}{t^2(1+t^2)}\ ,
\label{asymptote}
\end{equation}
where, $t\equiv\tan\theta$, with boundary conditions $\Psi(t=0)=0$,
and $\Psi(t=\infty)=\Psi_{\rm open}$. This yields good agreement with our
numerical simulation (fig.~3).

\section{The Outflow}

Up to this point we have said nothing about matter, except for a short
discussion of the space charge, which obviously consists of charged
matter particles. The reason is that the force--free problem that we
have attacked is complicated enough to also worry about the possible
effects of matter inertia and radiation reaction (charged particles
accelerated to relativistic velocities along curved trajectories emit
curvature radiation which can be thought of as an extra force acting
against the direction of motion of the particles). As is well known,
however, one can still consider a posteriori the motion of matter
along a force--free magnetic field geometry, provided the inertial and
radiation terms are assumed small enough compared to the electric and
magnetic terms which lead to the force--free solution
(Contopoulos~1995; Mestel \& Shibata~1994). In fact, the matter
problem {\em effectively decouples} from the electromagnetic one when
force--free conditions, flux freezing, and a cold plasma are assumed
throughout the flow. As is shown in great detail in the above
references, under those conditions, energy flux conservation implies
that
\begin{equation}
\gamma\left(1-\frac{R\Omega v_\phi}{c^2}\right)\ 
\label{gamma}
\end{equation}
is constant along field lines. Here, $\gamma\equiv (1-v^2/c^2)^{-1/2}$
is the Lorentz factor of the flow.  Moreover,
\begin{equation}
v_\phi=R\Omega+v_p\frac{B_\phi}{B_p}\ ,
\label{vphi}
\end{equation}
and therefore, knowing the distribution of the poloidal and
azimuthal field ($B_p$ and $B_\phi$ respectively) along a field line,
one can solve the system of equations~(\ref{gamma})--(\ref{vphi}), and
calculate the distribution of flow velocity everywhere (assuming an
initial outflow velocity distribution along the open field lines at
the surface of the star).

Before proceeding with the determination of the structure of the flow
along {\em the} solution obtained in the last section, one can make a
straightforward observation, namely that $\gamma$ would become
infinite whenever $v_\phi=c/x$ (we remind the reader that $x\equiv
R/R_{LC}$).  One can see that, through eq.~(\ref{vphi}), the latter
condition is equivalent to
\begin{equation}
\left|\frac{B_\phi}{B_p}\right|=\sqrt{x^2-1}\ ,
\label{infinity}
\end{equation}
and can only be satisfied outside the light cylinder.  In other words,
wherever eq.~(\ref{infinity}) is satisfied, $\gamma$ will diverge, and
the matter inertia cannot be considered as negligible anymore. Notice
that the condition in eq.~(\ref{infinity}) is a function of position
alone, along the force--free solution which we have
obtained. Obviously, a force--free solution knowing nothing about
matter inertia cannot, in general, require that
condition~(\ref{infinity}) be nowhere satisfied.  Interestingly
enough however, eq.~(\ref{infinity}) is satisfied nowhere along our solution
(at least within about 10 light cylinder radii from the origin, where
we trust the numerical accuracy of our solution most), which simply
implies that our force--free approximation is valid, even when matter
(with density not many orders of magnitude greater than the GJ
density) does flow along the field lines.  Note that $B_\phi/B_p<0$ and
thus $v_\phi< cx$, so $v_\phi$ will not approach too close to $c/x$,
as long as the ratio $|B_\phi/B_p|$ remains greater than
$\sqrt{x^2-1}$, as $x\rightarrow\infty$.

The interesting corollary of this discussion is that, if the Lorentz
factor of the outflow as it leaves the surface of the star is $\ll
10^6$, it does not approach the value of $10^6-10^7$, necessary to 
produce gamma ray emission by curvature radiation, anywhere near the light 
cylinder . This leads to the somewhat disappointing conclusion that
the magneto--centrifugal slingshot mechanism is not efficient in imparting
to the outflowing particles (electrons or positrons) the extremely 
relativistic energies required for the flow to generate gamma rays.  
A similar disappointing
conclusion has been reached for the problem of plasma acceleration in
a rotating relativistic split monopole (Bogovalov~1997). It has been
speculated in the above reference that the required acceleration may
result when one considers the time--dependent solution to the problem.
It is quite difficult to obtain the time--dependent solution of the
dipole problem because of the (awkward) boundary conditions of closed
and open field lines along the equator, and we tried instead to solve
the steady--state problem.  It appears, in retrospect, that our
expectations for an efficient (i.e. within $x \simeq $ a few)
acceleration of the flow to the required Lorentz factors ($10^6-10^7$)
by the rotating magnetic field pattern, especially near the equator,
have been overly optimistic.  If this is indeed the mechanism of
particle acceleration in pulsars, it will have to be achieved outside
the present framework of force--free, ideal magnetohydrodynamic
(${\bf E\cdot B} =0$) solutions.\footnote{More detailed analysis of
the flow properties (e.g. Shibata 1991; Mestel \& Shibata~1994) treats
the plasma as consisting of both the moderate--$\gamma$
electron--positron pairs that we considered above, and the
high--$\gamma$ primary electrons accelerated in the polar cap.}

We performed the straightforward exercise of solving the system of
eqs.~(\ref{gamma})--(\ref{vphi}), and the evolution of the flow
Lorentz factor along a representative field line is shown in
figure~5. We take $\gamma=10^3$ at the surface of the star (Daugherty
\& Harding~1982). It is very interesting that, although the Lorentz
factor initially slowly rises, it starts to decrease before the light
cylinder is crossed. This effect is unique to the dipolar geometry
that we have been investigating. As we have already mentioned, the
evolution of $\gamma$ is dictated by the evolution of the ratio
$|B_\phi/B_p|$ along the flow.  Interestingly enough, in the
relativistic monopole solution $|B_\phi/B_p|=x$ along all field (flow)
lines, and this implies a very gradual acceleration, as is seen in the
numerical simulation of Bogovalov~1997. In our present distorted
dipole solution, $B_\phi$ is again proportional to $1/x$, but $B_p$
decreases faster than $1/x^2$. This explains the flow {\em
deceleration} seen in our solution.

\section{Summary}

We have presented the first numerical solution of the structure of an
axisymmetric force--free magnetosphere due to an aligned magnetic
dipole under ideal MHD conditions; our solution joins smoothly
(i.e. without kinks/discontinuities) the (open) dipole field geometry,
interior to the light cylinder, to that of an outflowing MHD wind in
the asymptotic region. It would be very interesting if one could
complement our numerical results with an analytic solution, as is done
in the simple current free case inside the light cylinder (see Mestel
\& Pryce~1992 for a review). We were able to derive (semi--analytically) 
only the asymptotically radial structure of the flow.

The first important conclusion of our analysis is that we managed to
obtain numerically the unique (?) distribution of electric current
along the open field lines which is required in order for our present
magnetohydrodynamic model to be free of kinks/discontinuities near the
light cylinder. The open field lines from $\Psi=0$ to $1.36\Psi_{\rm
pc}$ are divided in 3 interesting parts: part (A) from $\Psi=0$ to
$\Psi_{\rm pc}$, where the main part of the electric current flows
into (out of, in a counter aligned rotator) the polar cap, without
crossing the null line; part (B) from $\Psi=1$ to $1.08\Psi_{\rm pc}$,
where electric current flows into (out of) the polar cap crossing the
null line twice; and part (C) from $\Psi=1.08$ to $1.36\Psi_{\rm pc}$,
where a small amount of return current flows out of (into) the polar
cap, crossing the null line once.  The electric circuit closes along
the boundary between closed and open field lines, and along the
equator.

We can tentatively speculate about the types of charge carriers in the
above electric currents. In part (A), the electric current will
naturally consist of {\em outflowing} electrons (positrons in a
counter aligned rotator) generated in a surface electrostatic gap. In
part (C), it will consist of {\em inflowing} electrons (positrons) and
{\em outflowing} positrons (electrons) generated in an outer gap along
the crossings of the null line.  Part (B) is more
complicated. Electron--positron pairs will be generated in the outer
crossings of the null line, and the electric current outside the inner
crossings of the null line will consist of {\em inflowing} positrons
(electrons) and {\em outflowing} electrons (positrons).  At the inner
crossings of the null line, the inflowing positrons (electrons) will
annihilate {\em outflowing} electrons (positrons) generated in a
surface electrostatic gap, and this completes the electric current
flow along part (B).  The return current along the boundary between
closed and open field lines, and along the equator, will consist of
equal amounts (to satisfy charge neutrality) of counter streaming flows
of electrons and positrons.  As we said, we plan to investigate the
detailed microphysics of these gaps and their associated high--energy
radiation processes in a forthcoming publication. We would like to
emphasize once again that the very existence of our smooth and
continuous MHD solution argues that nothing special happens at (or
about) the light cylinder.

The second important conclusion of our analysis is that, as in the
case of the relativistic monopole, the magnetocentrifugal slingshot
mechanism is not efficient in transforming the rotational 
spin--down energy flux carried by the electromagnetic field 
to particle relativistic kinetic energy.  Some other mechanism needs to be
invoked to account for the highly relativistic electrons necessary
to produce the observed pulsar $\gamma-$ray emission from the vicinity
of (or maybe well inside) the light cylinder. 
Magnetohydrodynamic flow instabilities
might be one interesting possibility, actively investigated by several
authors (e.g. Begelman~1998; Bogovalov~1997). Models with
$|B_\phi/B_p|\gg 1$ (as is the case in our model beyond the light
cylinder) tend to be unstable. Moreover, the magnetic field
discontinuity in the singular equatorial return current sheet might be
prone to reconnection, in which the current crosses the field by local
departure from the ideal MHD condition ${\bf E}\cdot {\bf B}=0$. Both
effects (assumed not to take place in our present idealized analysis)
will generate an effective large macroscopic resistivity. After all,
the breakdown of ideal MHD is inevitable, if we want to transform some
(significant) fraction of the electromagnetic field energy into  the
observable radiation from either the pulsar or the 
synchrotron nebula (plerion) which surrounds the pulsar
(Kennel \& Coroniti~1984). We believe that our present idealized
solution will help us better understand its origin.

\acknowledgements{We are grateful to our referee who helped us sketch 
a more physical picture of the pulsar magnetosphere.  We wish to
acknowledge interesting discussions with Sergei Bogovalov, Alice
Harding, Alexander Muslimov, Lev Titarchuk, Kanaris Tsinganos, and
Nektarios Vlahakis.  I.C. also wishes to acknowledge partial support
by NASA grants NAG 5-2266 and NAG 5-3687, and by grant 107526 from the
General Secretariat of Research and Technology of Greece.}

\newpage
\section*{Figure Captions}

{\bf Fig. 1.}--- Numerical checks of our integration routine.  In (a)
we run a simulation with $80\times 80$ points with AA'=0 inside the
light cylinder.  We plot the flux surfaces $\Psi=.15$, .4, 1.0, 1.4,
1.59 (heavy line), and $1.7\Psi_{\rm pc}$ respectively (we remind the
reader that $\Psi=0$ along the axis).  The solution compares well with
Michel~1991, fig.~4.9. In (b) we run a simulation with $30\times 30$
points inside the light cylinder and another $30\times 30$ points
outside for a rotating (split) monopole at the origin. The flux
$\Psi_{\rm m}=\Psi_{\rm open}(1-\cos\theta)$ and current $A_{\rm
m}=-R_{LC}^{-1}\Psi(2-\Psi/\Psi_{\rm open})$ distributions
respectively are obtained with high precision.  We plot the flux
surfaces $\Psi=.1$, .2, .3, .4, .5, .6, .7, .8, .9, and $1.0\Psi_{\rm
open}$ (heavy line).  In (c) we run a comparison with Michel~1991,
fig.~4.12, to show that, although the monopole current distribution
comes close to a smooth solution, it is not the final answer (a
numerical problem in that solution is discussed in the text). In that
simulation, we used the values of $\Psi(x=1^-,z)$ obtained from the
interior solution as boundary values $\Psi(x=1^+,z)$ in the exterior
solution.  Here, $A=-\Psi(2-\Psi/\Psi_{\rm open})$, and $\Psi_{\rm
open}=1.742\Psi_{\rm pc}$. We plot the flux surfaces $\Psi=.05$, .2,
.5, .9, 1.0, 1.1, and $1.5\Psi_{\rm open}$.

{\bf Fig. 2.}--- Evolution of the simulation for a rotating dipole at
the origin, as {\em the} correct current distribution is approached in
our iteration scheme ($1^{\rm st}$, $3^{\rm rd}$, $5^{\rm th}$
iteration in (a), (b), (c) respectively). Lines plotted as in fig.~3.

{\bf Fig. 3.}--- The final numerical solution for the structure of the
axisymmetric force--free magnetosphere of an aligned rotating magnetic
dipole.  We used a grid of $30\times 30$ points inside and another
$30\times 30$ points outside the light cylinder.  Thin lines represent
flux surfaces in intervals of $0.1\Psi_{\rm pc}$, with $\Psi=0$ along
the axis.  A small amount to return current flows between the dashed
field line $\Psi=1.08\Psi_{\rm pc}$ and the thick line at $\Psi_{\rm
open}=1.36\Psi_{\rm pc}$ which determines the boundary between closed
and open field lines, and where the bulk of the return current
flows. The null line, along which $\rho_e=0$, is shown dotted. The
solution asymptotes to the dashed--dotted lines obtained through the
integration of eq.~(\ref{asymptote}).

{\bf Fig. 4.}--- The electric current distribution $A=A(\Psi)$ (solid
line) along the open field lines which allows for the solution
presented in fig~3.  Compare this with the equivalent monopole (i.e. a
monopole with the same amount of open field lines) electric current
distribution $A_{\rm m}=-R_{LC}^{-1}\Psi(2-\Psi/\Psi_{\rm open})$
(dashed line).  Although our numerical iteration scheme seems to be
relaxing only to this unique distribution, we have no theoretical
arguments that this distribution is indeed unique.

{\bf Fig. 5.}--- The flow evolution along the dashed field line in
fig.~3. We plot the logarithm of the Lorentz factor $\gamma$ (solid),
$v_p$ (dashed), and $v_\phi$ (dotted). We took $\gamma=10^3$ at the
surface of the star.

\begin{figure}
\plotone{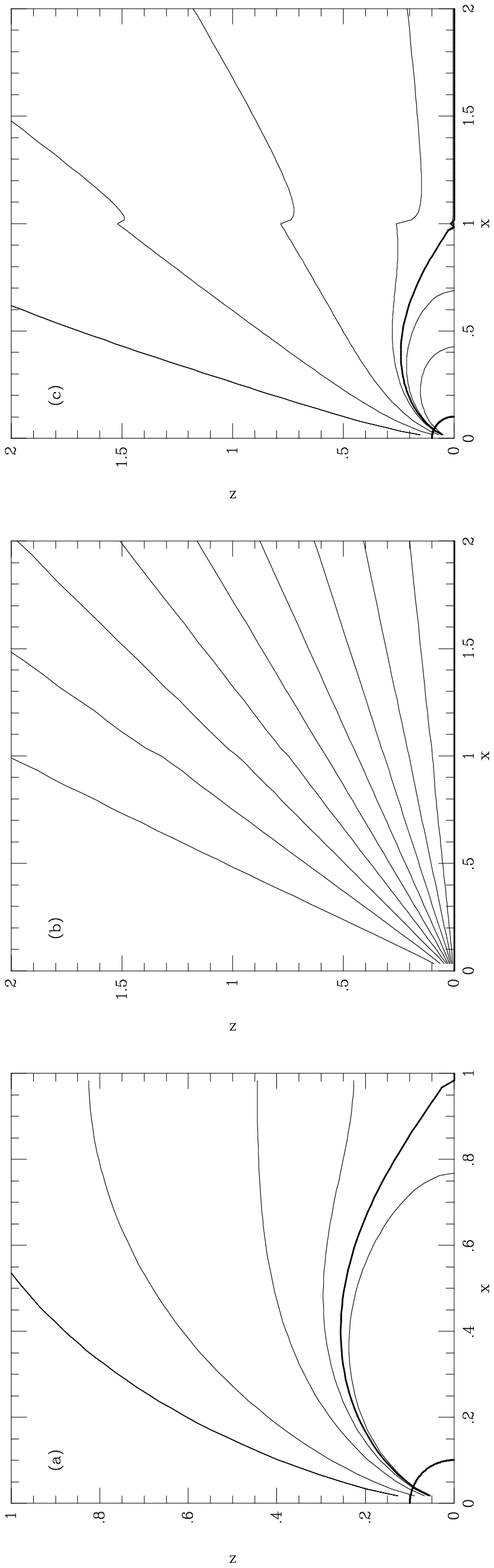}
\caption{}
\end{figure}

\begin{figure}
\plotone{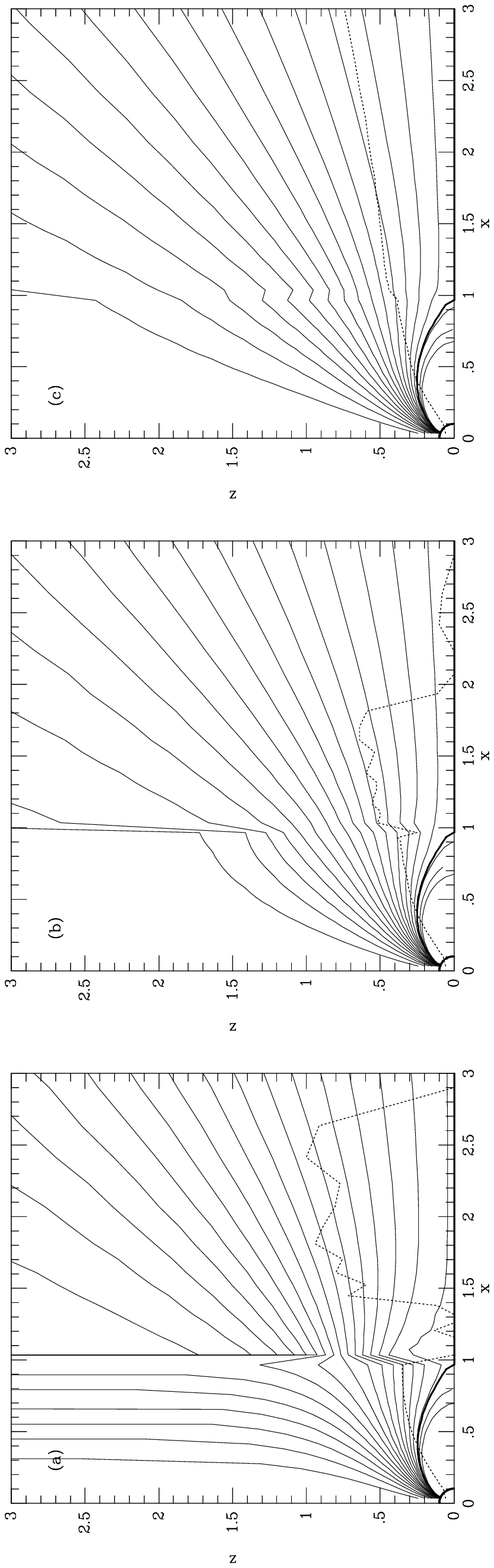}
\caption{}
\end{figure}

\begin{figure}
\plotone{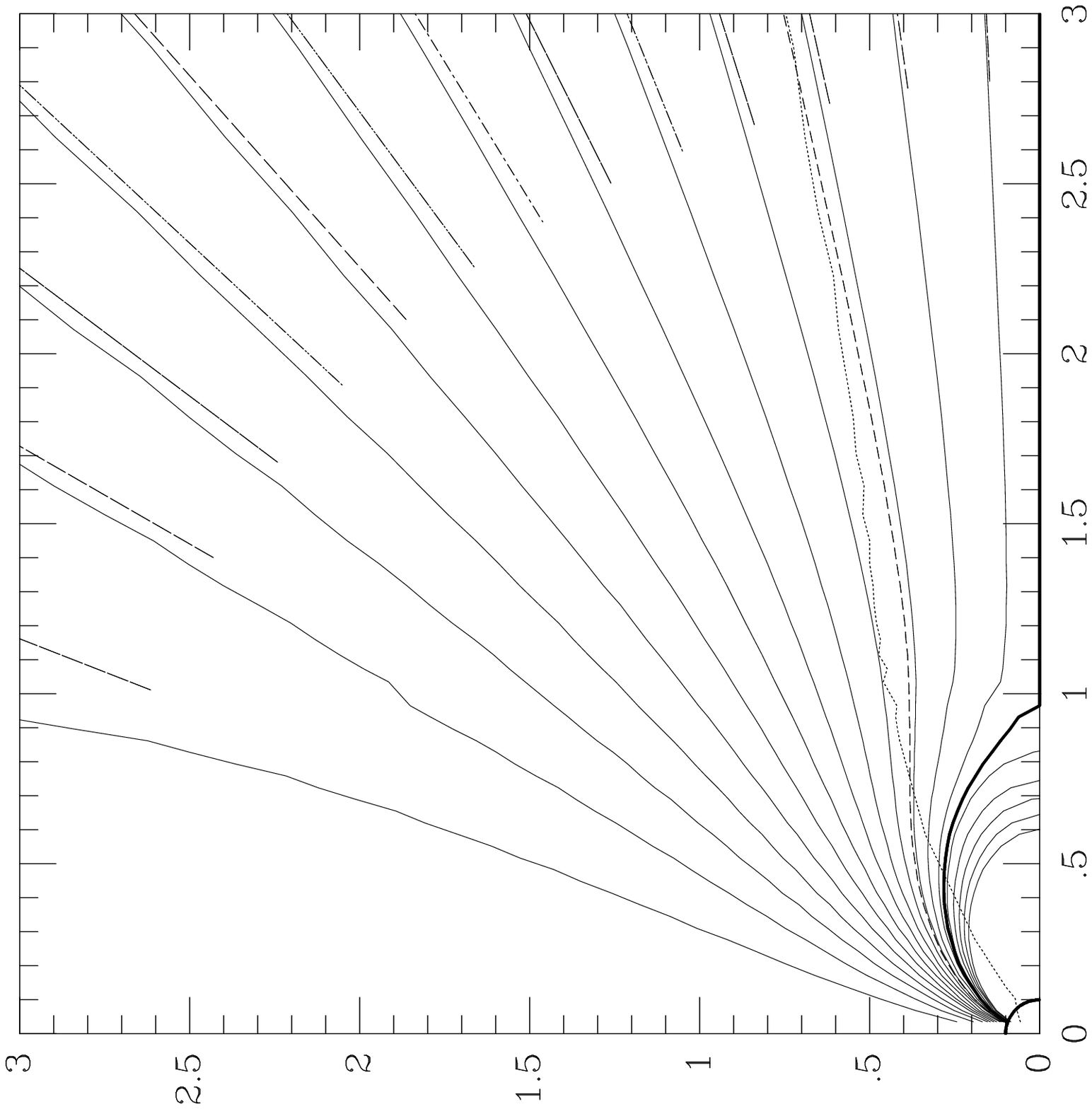}
\caption{}
\end{figure}

\begin{figure}
\plotone{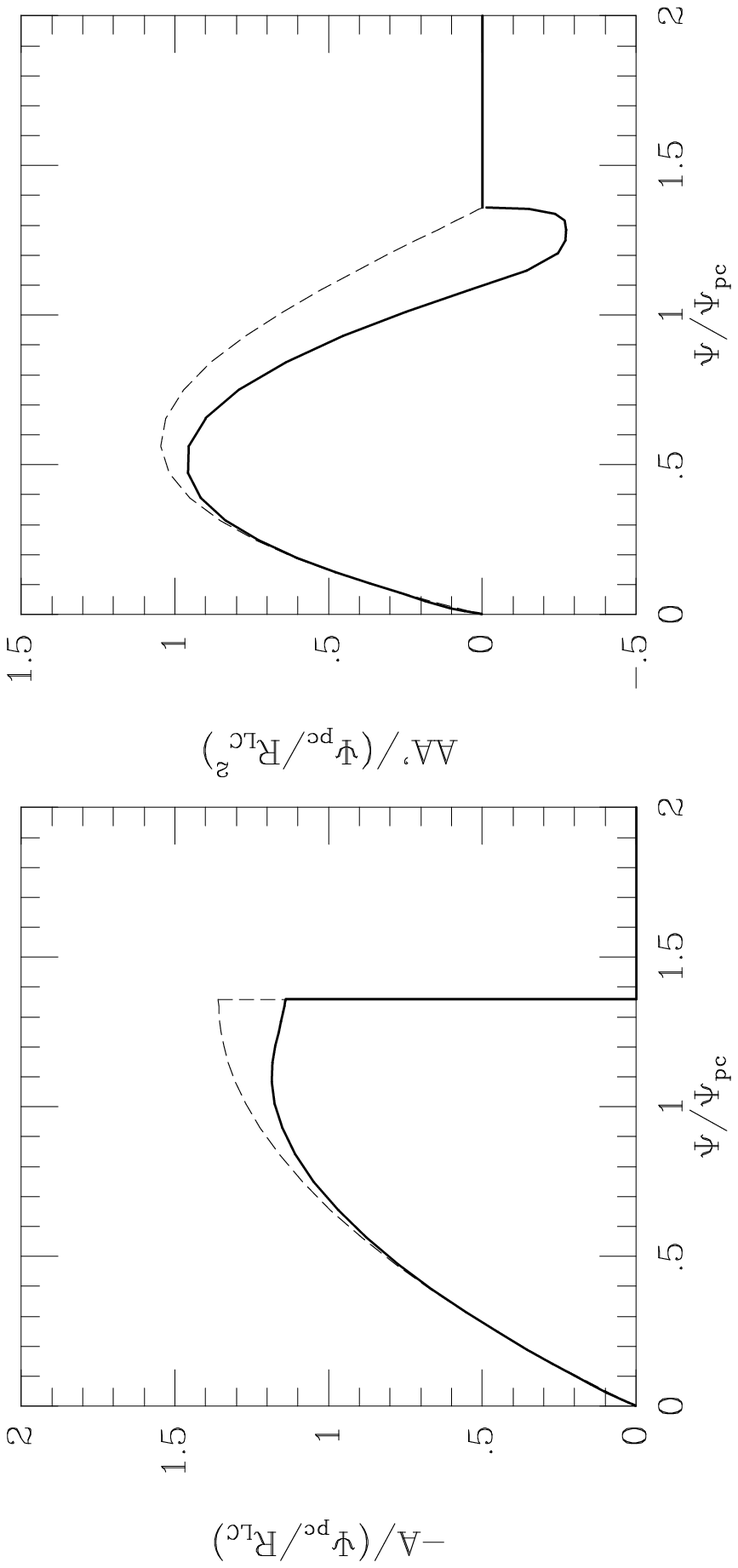}
\caption{}
\end{figure}

\begin{figure}
\plotone{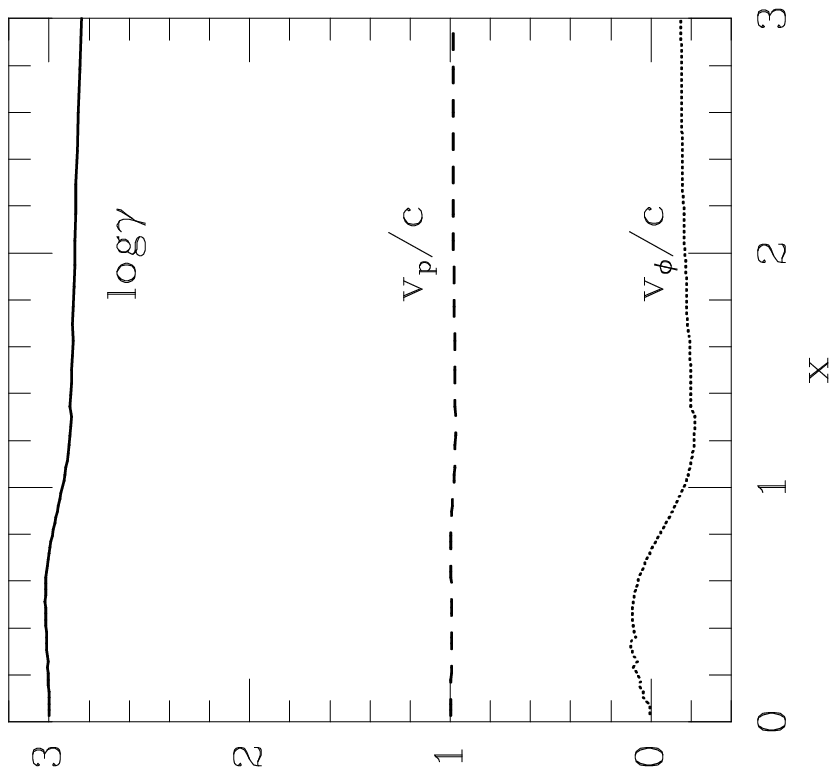}
\caption{}
\end{figure}

\end{document}